\RequirePackage{fixltx2e}
\documentclass[twocolumn,secnumarabic,superscriptaddress,amssymb, nobibnotes, aps, prb]{revtex4-1}
\usepackage{lmodern}
\usepackage{bbold}
\usepackage[figuresright]{rotating}
\usepackage[T1]{fontenc}
\usepackage{layouts}
\usepackage{amsfonts}
\usepackage{dcolumn}
\usepackage{amssymb}
\usepackage{amsmath}
\usepackage{graphicx}
\usepackage{supertabular}
\usepackage{bm}
\usepackage{bbm}
\usepackage{dsfont}
\usepackage{braket}
\usepackage{hyperref}
\usepackage{color,soul}
\usepackage[english]{babel}

\let\oldAA\AA
\renewcommand{\AA}{\text{\normalfont\oldAA}}
\newcommand{\m}{\mathcal}
\newcommand{\ofr}{(\bm{r})}
\newcommand{\ofSr}{(\{\m{S} \lvert \bm{f} \} \bm{r})}

\begin{document}

\title{ Density functional perturbation theory for lattice dynamics with fully relativistic ultrasoft pseudopotentials: the                                                                                    
magnetic case}%

\author{Andrea Urru}
\affiliation{International School for Advanced Studies (SISSA), \\
Via Bonomea 265, 34136 Trieste (Italy).}
\author{Andrea Dal Corso}
\affiliation{International School for Advanced Studies (SISSA), \\
Via Bonomea 265, 34136 Trieste (Italy).}
\affiliation{DEMOCRITOS IOM-CNR Trieste (Italy).}

\date{\today}%

\begin{abstract}
We extend density functional perturbation theory for lattice dynamics with fully 
relativistic ultrasoft pseudopotentials to magnetic materials.
Our approach is based on the application of the time-reversal operator to the Sternheimer
linear system and to its self-consistent solutions.
Moreover, we discuss how to include in the formalism the symmetry operations of the magnetic 
point group which require the time-reversal operator. We validate our implementation by comparison 
with the frozen phonon method in fcc Ni and in a monatomic ferromagnetic Pt wire.

\end{abstract}

\maketitle
%\tableofcontents
\section{Introduction}
Density Functional Perturbation Theory (DFPT) is widely used for the computation of the 
linear response properties of solids, and in particular for the study of their lattice dynamics. \cite{dfpt_review}
Some years ago, one of us applied DFPT \cite{us_fr_dfpt} to a scheme based on plane waves and norm conserving (NC) or 
ultrasoft (US) \cite{us_vanderbilt} pseudopotentials (PPs), that allow the introduction of spin-orbit effects within a 
fully relativistic (FR) density functional formalism and can be written in a form very similar to the scalar 
relativistic (SR) one. 
However, the theory presented in Ref. \onlinecite{us_fr_dfpt} was implemented only for time-reversal invariant systems, 
and therefore applications that include spin-orbit so far have been limited to non-magnetic solids. \cite{Pb_phonon} \\
In this work we extend this theory to the case of magnetic systems, by explicitly considering the presence 
of an exchange-correlation magnetic field in the Hamiltonian. DFPT equations in presence of a magnetic 
field have been recently written to calculate magnons with NC PPs in Refs. \onlinecite{Giustino} 
and \onlinecite{magnon}. In Ref. \onlinecite{Giustino}, the charge density induced by 
a periodic perturbation was computed by using the response to a perturbation at wave vector $\bm{q}$ and the response to 
a perturbation at $-\bm{q}$, 
while in Ref. \onlinecite{magnon} the problem at $-\bm{q}$ was not solved, but the time-reversal operator was used 
to obtain a second Sternheimer equation with a reversed magnetic field. The two formulations are equivalent. 
We generalize the theory of Ref. \onlinecite{magnon} to a phonon perturbation, avoiding the study of the 
response at $-\bm{q}$, and write it in a form applicable to both NC and US PPs. \\
In presence of a magnetic field, the solid is invariant upon the symmetry operations of the magnetic space 
group. Some of these operations require the time-reversal operator. We discuss how to exploit these symmetries 
for the symmetrization of the induced charge and magnetization densities and for the dynamical matrix. \\
Finally, we validate our method in ferromagnetic fcc Ni first computing the phonon frequencies at the $X$ point in 
the Brillouin zone (BZ) and comparing with DFPT results, and then by computing the phonon dispersions.  
Moreover, we apply our method to a monatomic ferromagnetic Pt nanowire and compare its 
vibrational properties when the magnetization is parallel or perpendicular to the wire. Also for this case 
we compare the DFPT results to the frozen phonon method for a phonon wavevector $q = \pi/a$ and $q = \pi/2a$, and then 
we compute by DFPT the phonon dispersion in the 1D BZ.

\section{DFPT with fully relativistic US-PPs}
In the Density Functional Theory (DFT) with FR US PPs, which accounts for spin-orbit effects, 
the minimization of the total energy functional leads to the Kohn-Sham \cite{KS} equations for the two-component spinor wave functions
\cite{us_fr_dfpt}: 
\begin{equation}
\sum_{\sigma '} H^{\sigma \sigma'} \ket{\Psi_i^{\sigma'}} = \epsilon_i \sum_{\sigma'} S^{\sigma \sigma'} \ket{\Psi_i^{\sigma'}},
\end{equation}
where $S^{\sigma \sigma'}$ is the overlap matrix needed in the US scheme, and the Hamiltonian $H^{\sigma \sigma'}$ is:
\begin{equation}
H^{\sigma \sigma'} = - \frac{1}{2} \nabla^2 \delta_{\sigma \sigma'} + V_{KS}^{\sigma \sigma'}.
\end{equation}
$V_{KS}^{\sigma \sigma'}$ is the total Kohn-Sham potential:
\begin{equation}
V_{KS}^{\sigma \sigma'} = V_{NL}^{\sigma \sigma'} + \sum_{\sigma_1 \sigma_2} \int d^3r V_{LOC}^{\sigma_1 \sigma_2} \ofr K_{\sigma \sigma'}^{\sigma_1 \sigma_2} \ofr,
\label{eq17}
\end{equation}
where: 
\begin{equation}
\begin{split}
K_{\sigma \sigma'}^{\sigma_1 \sigma_2} (\bm{r}, \bm{r}_1, \bm{r}_2) & = \delta(\bm{r} - \bm{r}_1) \, \delta(\bm{r} - \bm{r}_2) \, \delta_{\sigma_1 \sigma} \, \delta_{\sigma_2 \sigma'} \\ & 
+ \sum_{I m n} \sum_{m_1 n_1} Q_{m n}^{I} \ofr f_{m_1 m}^{\sigma \sigma_1} \beta_{m}^I (\bm{r}_1) \\ & \times f_{n n_1}^{\sigma_2 \sigma'} \beta_{n_1}^{* I} (\bm{r}_2),  
\end{split}
\label{eq18}
\end{equation} 
where $I = \{\rho, s''\}$, while $f_{m_1 m}^{\sigma_1 \sigma}$, $f_{n n_1}^{\sigma' \sigma_2}$, as well as the indeces $m$, 
$n$, $m_1$, and $n_1$ are defined in Eq. (5) of Ref. \onlinecite{us_fr_dfpt}.
In particular, in Eq. \eqref{eq17} $V_{LOC}^{\sigma \sigma'} = V_{eff} \delta_{\sigma \sigma'} - \mu_B \bm{B}_{xc} 
\cdot \bm{\sigma}^{\sigma \sigma'}$, and $V_{eff} = V_{loc} + V_H + V_{xc}$, is the sum of local, Hartree, and exchange and 
correlation potential, and $V_{NL}$ is the bare non local potential: in magnetic systems the spin, represented 
by the Pauli matrices, is coupled to the exchange-correlation magnetic field, 
$\bm{B}_{xc}$, defined as $\bm{B}_{xc} = -\delta E_{xc} / \delta \bm{m}$. This term breaks the time-reversal 
symmetry: indeed, introducing the time-reversal operator, $\m{T} = \imath \sigma_y \m{K}$, where 
$\m{K}$ is the complex-conjugation operator and $\sigma_y$ is the Pauli matrix, the following relationship holds:
\begin{equation}
\m{T} H^{[\bm{B}_{xc}]} \m{T}^{\dagger} = H^{[-\bm{B}_{xc}]}.
\label{eq1}
\end{equation}

We first exploit the time-reversal operator to rewrite the induced spin density. Following 
the notation of Ref. \onlinecite{us_fr_dfpt}, we consider both the metallic and the insulating case. The change of the 
spin density induced by the variation of an external parameter $\mu$ (Eq. (10) of Ref. \onlinecite{us_fr_dfpt}) may 
be written as:
\begin{equation}
\begin{split}
\frac{dn^{\sigma \sigma'} \ofr}{d \mu} = & \sum_{i} \sum_{\sigma_1 \sigma_2} \bigg[ \bra{\Psi_i^{\sigma_1}} 
K_{\sigma_1 \sigma_2}^{\sigma \sigma'} \ofr \ket{\Delta^{\mu} \Psi_i^{\sigma_2}} \\ 
& + \sum_{\sigma_3 \sigma_4} \bra{\left(\m{T} \Psi_i\right)^{\sigma_1}} \m{T}_{\sigma_1 \sigma_3} \, K_{\sigma_3 \sigma_4}^{\sigma' \sigma} \ofr \, \m{T}_{\sigma_4 \sigma_2}^{\dagger} \\ 
& \times \ket{\left( \m{T} 
\Delta^{\mu} \Psi_i \right)^{\sigma_2}} \bigg] + \Delta^{\mu} n^{\sigma \sigma'} \ofr.
\end{split}
\label{eq2}
\end{equation}
$\Delta^{\mu} n^{\sigma \sigma'}$ is defined as in the non-magnetic case and corresponds to 
the last two terms of Eq. (10) of Ref. \onlinecite{us_fr_dfpt}.

The same idea can be applied to the second-order derivatives of the total energy. Only the term 
$d^2 E_{tot}^{(2)}/d \mu d\lambda$ (Eq. (19) of Ref. \onlinecite{us_fr_dfpt}) needs to be rewritten 
by using $\m{T}$:
%
%\begin{widetext}
\begin{equation}
\begin{split}
& \frac{d^2 E_{tot}^{(2)}}{d \mu d\lambda} = \sum_{i} \sum_{\sigma_1 \sigma_2} \langle \Psi_i^{\sigma_1} \Biggm\lvert \frac{\partial V_{KS}^{[\bm{B}] 
\sigma_1 \sigma_2}}{\partial \lambda} - \epsilon_i \frac{\partial S^{\sigma_1 \sigma_2}}{\partial \lambda} \Biggm\lvert
\Delta^{\mu} \Psi_i^{\sigma_2} \rangle \\ 
& + \sum_{i} \sum_{\sigma_1 \sigma_2} \langle \left(\m{T} \Psi_i\right)^{\sigma_1}
\Biggm\lvert \frac{\partial V_{KS}^{[-\bm{B}] \sigma_1 \sigma_2}}{\partial \lambda} - \epsilon_i \frac{\partial S^{\sigma_1 \sigma_2}}
{\partial \lambda} \Biggm\lvert \left( \m{T} \Delta^{\mu} \Psi_i \right)^{\sigma_2} \rangle,
\end{split}
\label{eq3}
\end{equation}
%\end{widetext}
%
while the other contributions can be kept in their original form. Both Eq. \eqref{eq2} and Eq. \eqref{eq3} contain 
two unknown terms, namely $\ket{\Delta^{\mu} \Psi_i^{\sigma}}$ and $\ket{(\m{T} \Delta^{\mu} \Psi_i)^{\sigma}}$. 
The first can be computed by means of the Sternheimer linear system (Eqs. 13 and 14 of Ref. \onlinecite{us_fr_dfpt}), 
while the second is obtained by solving the following linear system:
\begin{equation}
\begin{split}
& \sum_{\sigma_2} \left[ H^{[-\mathbf{B}] \sigma_1 \sigma_2} - \epsilon_i S^{\sigma_1 \sigma_2} \right] \Bigm \lvert \left( \m{T} 
\Delta^{\mu} \Psi_i \right)^{\sigma_2} \rangle = \\ & - \sum_{\sigma_2 \sigma_3} \Pi_{c,i}^{\dagger \sigma_1 \sigma_2} \left[ 
\frac{dV_{KS}^{[-\mathbf{B}] \sigma_2 \sigma_3}}{d \mu} - \epsilon_i \frac{\partial S^{\sigma_2 \sigma_3}}{\partial \mu} \right]
\Biggm \lvert \left(\m{T} \Psi_i\right)^{\sigma_3} \rangle, 
\end{split}
\label{eq4}
\end{equation}
obtained by applying $\m{T}$ to both sides of the Sternheimer linear system (Eq. (13) of Ref. \onlinecite{us_fr_dfpt}) and 
using the fact that $\m{T} \, dV_{KS}^{[\bm{B}]}/d \mu \, \m{T}^{\dagger} = dV_{KS}^{[-\bm{B}]}/d \mu$. In particular, here we 
introduced the time-reversed projector on the conduction manifold, namely 
$\Pi_{c,i}^{\dagger \sigma_1 \sigma_2} = \sum_{\sigma \sigma'} \m{T}_{\sigma_1 \sigma} P_{c,i}^{\dagger \sigma \sigma'} \m{T}^{\dagger}_{\sigma' \sigma_2}$, 
similarly to what proposed in Ref. \onlinecite{magnon} for the calculation of magnons.
Eqs. \eqref{eq2}, \eqref{eq3}, and \eqref{eq4} are valid for the US PPs scheme, the 
NC formulation can be obtained as a particular case by writing 
$K_{\sigma_1 \sigma_2}^{\sigma \sigma'}(\bm{r},\bm{r}_1,\bm{r}_2) = \delta(\bm{r} - \bm{r}_1) \, \delta(\bm{r} - \bm{r}_2) \, \delta_{\sigma \sigma_1} \, \delta_{\sigma' \sigma_2}$ 
and $S^{\sigma_1 \sigma_2} = \delta_{\sigma_1 \sigma_2}$. Moreover, the insulating case can be dealt with by putting 
$\widetilde{\theta}_{F,i} = 1$ if the state is occupied or $0$ if the state is empty (see Ref. \onlinecite{us_dfpt} for the 
definition of $\widetilde{\theta}_{F,i}$).

\section{Phonons in periodic solids}
In this section, we consider a phonon perturbation with a wavevector $\bm{q}$ perturbing a periodic solid, for which 
the wave functions $\Psi_i^{\sigma} \ofr$ may be written in the Bloch form, $\Psi_{\bm{k} v}^{\sigma} \ofr= e^{\imath 
\bm{k} \cdot \bm{r}} u_{\bm{k} v}^{\sigma} \ofr$, where $u_{\bm{k} v}^{\sigma} \ofr$ is lattice periodic. 
Following the discussion reported in Appendix A of Ref. \onlinecite{us_dfpt}, we introduce the variation of the density 
and of the wave functions, induced by a phonon perturbation $u_{\nu s' \beta} = 1/\sqrt{M_{s'}} \Re \left( u_{s' \beta} 
(\bm{q}) e^{\imath \bm{q} \cdot \bm{R_{\nu}}} \right)$, where $\Re$ indicates the real part, and define:
\begin{align}
\frac{dn^{\sigma \sigma'} \ofr}{d u_{s' \beta} (\bm{q})} & = \sum_{\nu} e^{\imath \bm{q} \cdot \bm{R_{\nu}}} \frac{dn^{\sigma \sigma'} \ofr}{d u_{\nu s' \beta}}, \\
\ket{\Delta^{u_{s' \beta} (\bm{q})} \Psi_{\bm{k}v}^{\sigma}} & = \sum_{\nu} e^{\imath \bm{q} \cdot \bm{R_{\nu}}} \ket{\Delta^{u_{\nu s' \beta}} \Psi_{\bm{k}v}^{\sigma}}.
\label{eq14}
\end{align}
Eq. \eqref{eq2} then becomes:  
\begin{equation}
\begin{split}
\frac{dn^{\sigma \sigma'} \ofr}{d u_{s' \beta} (\bm{q})} & = \sum_{\bm{k} v} \sum_{\sigma_1 \sigma_2} \bigg[ \bra{\Psi_{\bm{k} v}^{\sigma_1}} K_{\sigma_1 \sigma_2}^{\sigma \sigma'} \ofr \ket{\Delta^{u_{s' \beta} (\bm{q})} \Psi_{\bm{k} v}^{\sigma_2}} \\ & + \sum_{\sigma_3 \sigma_4} \bra{\left(\m{T} \Psi_{- \bm{k} v}\right)^{\sigma_1}} T_{\sigma_1 \sigma_3} K_{\sigma_3 \sigma_4}^{\sigma' \sigma} \ofr T_{\sigma_4 \sigma_2}^{\dagger} \\ 
& \times \ket{\left( \m{T} \Delta^{u_{s' \beta} (- \bm{q})} \Psi_{-\bm{k} v} \right)^{\sigma_2}} \bigg] + \Delta^{u_{s' \beta} (\bm{q})} n^{\sigma \sigma'} \ofr. 
\end{split}
\label{eq10}
\end{equation}
In Eq. \eqref{eq10} the second term is identical to the first and is not explicitly computed in time-reversal invariant systems. The same holds for 
\eqref{eq11} below (for the dynamical matrix). Instead, for magnetic systems the two terms must be computed separately. In particular, the 
time-reversed response of the wave functions can be computed by solving the linear system Eq. \eqref{eq4}, which becomes: 
\begin{equation}
\begin{split}
& \sum_{\sigma_2} \left[ H^{[-\bm{B}] \sigma_1 \sigma_2} - \epsilon_{-\bm{k} v} S^{\sigma_1 \sigma_2} \right] \ket{(\m{T} \Delta^{u_{s' \beta} (- \bm{q})} \Psi_{- \bm{k} v})^{\sigma_2}} = \\
& - \sum_{\sigma_2} \Pi_{c, -\bm{k}v}^{\dagger \sigma_1 \sigma_2} \Bigg[ \ket{\phi_{T -\bm{k}v}^{u_{s' \beta} (\bm{q}) [-\bm{B}]\sigma_2}} \\ & + \sum_{\sigma_3} \sum_{\sigma_4 \sigma_5} \int d^3 r \frac{d V_{H,xc}^{[-\bm{B}] \sigma_4 \sigma_5} \ofr }{d u_{s' \beta} (\bm{q})} K^{\sigma_4 \sigma_5}_{\sigma_2 \sigma_3} (\bm{r}) \ket{\left(\m{T} \Psi_{- \bm{k} v}\right)^{\sigma_3}} \Bigg],
\end{split}
\label{eq13}
\end{equation}
where, similarly to Ref. \onlinecite{us_dfpt} we defined:
\begin{equation}
\ket{\phi_{T -\bm{k} v}^{u_{s' \beta} (\bm{q}) [-\bm{B}] \sigma_2}} = \sum_{\sigma_3} \left( \frac{\partial V_{KS}^{[-\bm{B}] \sigma_2 \sigma_3}}{\partial u_{s' \beta} (\bm{q})} - \epsilon_{-\bm{k} v} \frac{\partial S^{\sigma_2 \sigma_3}}{\partial u_{s' \beta} (\bm{q})} \right) \ket{\left(\m{T} \Psi_{-\bm{k} v}\right)^{\sigma_3}},
\end{equation}
in which $ \partial V_{KS}^{[-\bm{B}] \sigma_2 \sigma_3} / \partial u_{s' \beta} (\bm{q})$ and $\partial S^{\sigma_2 \sigma_3} / \partial u_{s' \beta} (\bm{q})$
are defined similarly to Eq. \eqref{eq14}. The action of the time-reversal operator on the linear system changes the sign of the exchange 
and correlation magnetic field, which enters in the Hamiltonian, in $d V_{H, xc}^{[\bm{B}] \sigma_4 \sigma_5} / d u_{s' \beta} (\bm{q})$, 
and in $\ket{\phi^{u_{s' \beta} (\bm{q}) [\bm{B}] \sigma_2}_{\bm{k} v}}$ through the third term in Eq. (9) of Ref. \onlinecite{us_fr_dfpt}.
We can then write the contribution to the dynamical matrix coming from $d^2 E_{tot}^{(2)}/d u_{\mu s \alpha} d u_{\nu s' \beta}$ 
in the following way:
\begin{equation}
\begin{split}
\Phi^{(2)}_{\substack{s \alpha \\ s' \beta}} (\bm{q}) & = \frac{1}{N} \sum_{\bm{k} v} \sum_{\sigma} \Bigg[ \langle \phi^{u_{s \alpha} (\bm{q}) [\bm{B}] \sigma}_{\bm{k} v} \lvert \Delta^{u_{s' \beta} (\bm{q})} \Psi_{\bm{k} v}^{\sigma} \rangle \\ & + \langle \phi^{u_{s \alpha} (\bm{q}) [-\bm{B}] \sigma}_{T -\bm{k} v} \lvert \left( \m{T} \Delta^{u_{s' \beta} (-\bm{q})} \Psi_{-\bm{k} v} \right) ^{\sigma} \rangle \Bigg],
\end{split}
\label{eq11}
\end{equation}
while the other contributions may be kept in their original form, discussed in Ref. \onlinecite{us_dfpt}. Here, $N$ is the 
number of cells in the solid.

Eq. \eqref{eq10} may be further manipulated by writing explicitly $K_{\sigma_1 \sigma_2}^{\sigma \sigma'} \ofr$ 
(Eq. \eqref{eq18}). Introducing the periodic parts of the Bloch functions and of the responses of the wave functions, 
we obtain the periodic part of the induced spin density (indicated with a tilde ($\sim$)):
\begin{equation}
\begin{split}
\frac{{\widetilde{dn^{\sigma \sigma'}} \ofr}}{d u_{s' \beta} (\bm{q})} & = \sum_{\bm{k} v} \Bigg[ u_{\bm{k} v}^{* \, \sigma}\ofr \widetilde{\Delta^{u_{s' \beta} (\bm{q})}} u_{\bm{k} v}^{\sigma'} \ofr + \sum_{\sigma_1 \sigma_2} U_{\sigma' \sigma_1} \\ & \times (\m{T} u_{- \bm{k} v} \ofr)^{* \, \sigma_1} (\m{T} \widetilde{\Delta^{u_{s' \beta} (- \bm{q})}} u_{- \bm{k} v} \ofr)^{\sigma_2} U_{\sigma_2 \sigma}^{\dagger} \Bigg] \\ & + \sum_{s'' m \, n} \bigg( \widetilde{Q}_{m n}^{s'' \bm{q}} \ofr \Delta^{u_{s' \beta} (\bm{q})} \rho_{m n}^{s'' \sigma \sigma'}\bigg) \\ 
& + \widetilde{\Delta^{u_{s' \beta} (\bm{q})}} n^{\sigma \sigma'} \ofr,
\end{split}
\label{eq12}
\end{equation}
where we defined the quantities $\widetilde{Q}_{m n}^{s'' \bm{q}} \ofr$ and $\Delta^{u_{s' \beta} (\bm{q})} \rho_{m n}^{s'' \sigma \sigma'}$ as: 
\begin{equation}
\widetilde{Q}_{m n}^{s'' \bm{q}} \ofr = e^{-\imath \bm{q} \cdot \bm{r}} \sum_{\rho} e^{\imath \bm{q} \cdot \bm{R}_{\rho}} Q_{m n}^{I} \ofr ,
\end{equation}
\begin{widetext}
\begin{equation}
\begin{split}
\Delta^{u_{s' \beta} (\bm{q})} \rho_{m n}^{s'' \sigma \sigma'}  & = \sum_{m_1 n_1} \sum_{\sigma_1 \sigma_2} \sum_{\bm{k} v} \Big( \beta_{\bm{k} v}^{* \, s'' m_1 \sigma_1 } f_{m_1 m}^{\sigma_1 \sigma} f_{n n_1}^{\sigma' \sigma_2} \Delta^{u_{s' \beta} (\bm{q})} \beta_{\bm{k} v}^{s'' n_1 \sigma_2} \\ & + \sum_{\sigma'' \sigma'''} \beta_{T \, -\bm{k} v}^{* \, s'' m_1 \sigma_1} f_{m_1 m}^{\sigma_1 \sigma''} U_{\sigma'' \sigma'} U_{\sigma \sigma'''}^{\dagger} f_{n n_1}^{\sigma''' \sigma_2}
\Delta^{u_{s' \beta} (- \bm{q})} \beta_{T \, -\bm{k} v}^{s'' n_1 \sigma_2} \Big),
\end{split}
\label{eq15}
\end{equation}
\end{widetext}
where $\beta_{T \, -\bm{k} v}^{s'' m_1 \sigma_1} = e^{- \imath \bm{k} \cdot \bm{R}_{\rho}} \, \langle \beta^{I}_{m_1} \lvert 
(\m{T} \Psi_{- \bm{k} v})^{\sigma_1}\rangle$, $\Delta^{u_{s' \beta} (- \bm{q})} \beta_{T \, -\bm{k} v}^{s'' n_1 \sigma_2} = 
e^{- \imath (\bm{k}+\bm{q}) \cdot \bm{R}_{\rho}} \langle \beta^{I}_{n_1} \lvert (\m{T} \Delta^{u_{s' \beta} (- \bm{q})} \Psi_{- \bm{k} v})^{\sigma_2} \rangle$, 
and we used the fact that: 
\begin{equation}
\sum_{\sigma_3 \sigma_4} \m{T}_{\sigma_1 \sigma_3} K_{\sigma_3 \sigma_4}^{\sigma' \sigma} \ofr \m{T}_{\sigma_4 \sigma_2}^{\dagger} = \sum_{\sigma'' \sigma'''} U_{\sigma' \sigma''} K_{\sigma_1 \sigma_2}^{\sigma'' \sigma'''} U_{\sigma''' \sigma}^{\dagger},
\end{equation}
where $U = \imath \sigma_y$ is the unitary part of the time-reversal operator.
The induced charge and magnetization densities can be computed from the induced spin density (Eq. \eqref{eq12}) as: 
\begin{align}
\frac{\widetilde{d n}\ofr}{d u_{s \alpha}(\bm{q})} & = \sum_{\sigma} \frac{{\widetilde{dn^{\sigma \sigma}} \ofr}}{d u_{s \alpha} (\bm{q})}, \\
\frac{\widetilde{d m}_{\beta}\ofr}{d u_{s \alpha}(\bm{q})} & = \mu_B \sum_{\sigma \, \sigma'} \frac{{\widetilde{dn^{\sigma \sigma'}} \ofr}}{d u_{s \alpha} (\bm{q})} \sigma_{\beta}^{\sigma \sigma'}.
\label{eq16}
\end{align}
In particular, for the induced charge density we use the fact that $\sum_{\sigma} U_{\sigma_2 \sigma}^{\dagger} U_{\sigma \sigma_1} = \delta_{\sigma_1 \sigma_2}$ 
and for the induced magnetization density the fact that 
$\sum_{\sigma \sigma'} U_{\sigma_2 \sigma}^{\dagger} \sigma_{\alpha}^{\sigma \sigma'} U_{\sigma' \sigma_1} = - \sigma_{\alpha}^{\sigma_1 \sigma_2}$, 
so that the terms of the induced spin density that contain the time-reversed wave functions are subtracted in Eq. \eqref{eq16}.

The linear system \eqref{eq13} may be written in terms of lattice periodic functions: 
\begin{equation}
\begin{split}
& \sum_{\sigma_2} \left( {H_{\bm{k} + \bm{q}}^{[-\bm{B}] \sigma_1 \sigma_2}} - \epsilon_{-\bm{k} v} S_{\bm{k} + \bm{q}}^{\sigma_1 \sigma_2}\right) \ket{(\m{T} \widetilde{\Delta^{u_{s' \beta} (- \bm{q})}} u_{- \bm{k} v})^{\sigma_2}} \\ & = - \sum_{\sigma_2} \Pi_{c, -\bm{k} v}^{\dagger \sigma_1 \sigma_2, -\bm{k} - \bm{q}} \Bigg[ \ket{\widetilde{\phi}_{T - \bm{k} v}^{u_{s' \beta} (\bm{q}) [-\bm{B}] \sigma_2}} + \sum_{\sigma_3} \frac{\widetilde{d V}_{H, xc}^{[- \bm{B}] \sigma_2 \sigma_3}}{d u_{s' \beta} (\bm{q})} \\ & \times \ket{(\m{T} u_{\bm{k} v})^{\sigma_3}} + \sum_{\sigma_3} \sum_{s'' m_1 n_1} {}^{3} I_{s'' m_1 n_1}^{u_{s' \beta} (\bm{q}) [-\bm{B}] \sigma_2 \sigma_3} \ket{\widetilde{\beta}_{m_1 \bm{k} + \bm{q}}^{s''}} \\ & \times \beta_{T \, -\bm{k} v}^{s'' n_1 \sigma_3} \Bigg],
\end{split}
\end{equation}
where we defined: 
\begin{equation}
\langle \bm{r} \ket{\widetilde{\beta}_{m_1 \bm{k} + \bm{q}}^{s''}} = e^{-\imath (\bm{k} + \bm{q}) \cdot \bm{r}} \sum_{\rho} e^{\imath (\bm{k} + \bm{q}) \cdot \bm{R}_{\rho}} \beta_{m_1}^{s''} (\bm{r} - \bm{R}_I)
\end{equation}
\begin{equation}
{}^{3} I_{s'' m_1 n_1}^{u_{s' \beta} (\bm{q}) [-\bm{B}] \sigma_2 \sigma_3} = \sum_{m n} \sum_{\sigma_4 \sigma_5} f_{m_1 m}^{\sigma_2 \sigma_4} f_{n n_1}^{\sigma_5 \sigma_3} \sum_{\alpha} A_{\alpha}^{\sigma_4 \sigma_5} {}^{3} I_{s'' m n \alpha}^{u_{s' \beta} (\bm{q}) [-\bm{B}]},
\end{equation}
and:
\begin{equation}
{}^{3} I_{s'' m n \alpha}^{u_{s' \beta} (\bm{q}) [-\bm{B}]} = \int Q_{m n}^{\gamma(s'')} (\bm{r} - \bm{d}_{s''}) \, \frac{d C_{\alpha}^{[-\bm{B}]} \ofr}{d u_{s' \beta} (\bm{q})} \, d^3 r, 
\end{equation}
where $\alpha = 1, \dots, 4$, $\bm{A} = (\mathbb{1}, \sigma_x, \sigma_y, \sigma_z)$, similarly to Ref. \onlinecite{us_fr_pseudo} 
($\mathbb{1}$ is the $2 \times 2$ identity matrix), and $\bm{C}^{[\bm{B}]} = (V_{H,xc}, -\mu_B B_{xc, x}, -\mu_B B_{xc, y}, -\mu_B B_{xc, z})$. 

\section{symmetrization}
\label{sec4}
We indicate with $\{\m{S} \lvert \bm{f}\}$ the symmetry operations of the space group of the crystal, where $\m{S}$ 
is a rotation (proper or improper) and $\bm{f}$ is a translation. In a magnetic crystal, we have to 
consider also the operations $\m{S}$ such that $\{\m{T} \m{S} \lvert \bm{f} \}$ is a symmetry of the crystal.

Since, for a phonon perturbation, the charge (and magnetization) density response and the dynamical matrix 
are computed at a given finite wave vector $\bm{q}$, we use as symmetry operations only those $N_S$ operations 
of the antiunitary small space group of $\bm{q}$, the subgroup of the antiunitary space group of the crystal, which 
contains the symmetry operations $\{\m{S} \lvert \bm{f} \}$ such that:
\begin{equation}
\m{S} \bm{q} = \bm{q} + \bm{G}_{\m{S}},
\label{eq5}
\end{equation}
if $\{\m{S} \lvert \bm{f} \}$ is a symmetry of the crystal, or:
\begin{equation}
\m{S} \bm{q} = - \bm{q} + \bm{G}_{\m{S}},
\label{eq6}
\end{equation}
if $\{ \m{T} \m{S} \lvert \bm{f} \}$ is a symmetry of the crystal. Here, $\bm{G}_{\m{S}}$ is a reciprocal lattice vector 
that might appear when $\bm{q}$ is at zone border. In order to distinguish the two cases we introduce a variable 
$\tau (\m{S})$ which may take the values $\tau = 0$ or $\tau = 1$ if Eq. \eqref{eq5} or Eq. \eqref{eq6} holds, respectively.
We compute the unsymmetrized induced spin density by summing over the Irreducible Brillouin Zone (IBZ) in Eqs. 
\eqref{eq12} and \eqref{eq15}, introducing a weight proportional to the number of elements in the star of $\bm{q}$. 
Then, we calculate the unsymmetrized induced charge and magnetization densities $\widetilde{d n}^{NS}\ofr/d u_{s' \beta}(\bm{q})$ 
and $\widetilde{d m}_{\delta}^{NS}\ofr/d u_{s' \beta}(\bm{q})$ using Eqs. (19) and \eqref{eq16}. Finally, the complete 
responses are obtained through the following relationships:
\begin{equation}
\begin{split}
\frac{\widetilde{d n \ofr}}{d u_{s' \beta}(\bm{q})} =  \frac{1}{N_S} \sum_{\{\m{S} \lvert \bm{f} \}} \m{O}_{\tau(\m{S})} & \bigg[ \sum_{\gamma} S_{\gamma \beta} \frac{\widetilde{d n}^{NS}\ofSr}{d u_{\bar{s'} \gamma}(\bm{q})} \\ & \times e^{\imath \bm{G}_{\m{S}^{-1}} \cdot \bm{r}} e^{- \imath \bm{q} \cdot \bm{R}^{\m{S}}_{s'}} \bigg], 
\end{split}
\label{eq7}
\end{equation}
\begin{equation}
\begin{split}
\frac{\widetilde{d m_{\delta} \ofr}}{d u_{s' \beta}(\bm{q})} & =  \frac{1}{N_S} \sum_{\{\m{S} \lvert \bm{f} \}} (-1)^{\tau(\m{S})} \m{O}_{\tau(\m{S})} \Bigg[ \sum_{\gamma \eta} \tilde{S}^{-1}_{\delta \eta} S_{\gamma \beta} \\ & \times \frac{\widetilde{d m}_{\eta}^{NS}\ofSr}{d u_{\bar{s'} \gamma}(\bm{q})} e^{\imath \bm{G}_{\m{S}^{-1}} \cdot \bm{r}} e^{- \imath \bm{q} \cdot \bm{R}^{\m{S}}_{s'}} \Bigg],
\end{split}
\label{eq8}
\end{equation}
where $\tilde{S}$ is the proper part of $S$, $\m{O}_{\tau(\m{S})}$ is the identity if $\tau(\m{S})=0$, 
or $\m{O}_{\tau(\m{S})} = \m{K}$ if $\tau(\m{S})=1$. Moreover, $\bm{R}^{\m{S}}_{s'} = 
\m{S} \bm{d}_{s'} - \bm{d}_{\bar{s'}}$, where $\bm{d}_{s'}$ identifies the position of the atom $s'$ with 
respect to the origin of its primitive cell, while $\bm{d}_{\bar{s'}}$ is obtained by applying the 
rotation $\m{S}$ to the atom $s'$ ($\{\m{S} \lvert \bm{f} \} (\bm{R}_{\nu} + \bm{d}_{s'}) = \bm{R}_{\bar \nu} + \bm{d}_{\bar{s'}}$). 
Similarly, the dynamical matrix becomes:
\begin{equation}
\begin{split}
\Phi_{\substack{s \alpha \\ s' \beta}} (\bm{q}) = \frac{1}{N_S} \sum_{\{\m{S} \lvert \bm{f} \}} \m{O}_{\tau(\m{S})} & \Bigg[ \sum_{\gamma \delta} S_{\gamma \alpha} \, S_{\delta \beta} \, \Phi_{\substack{\bar{s} \gamma \\ \bar{s'} \delta}}^{NS} (\bm{q}) \\ & \times e^{\imath \bm{q} \cdot (\bm{R}^{\m{S}}_{s} - \bm{R}^{\m{S}}_{s'})} \Bigg],
\end{split}
\label{eq9}
\end{equation}
where $\Phi_{\substack{\bar{s} \gamma \\ \bar{s'} \delta}}^{NS} (\bm{q})$ is obtained summing over the IBZ in 
Eq. \eqref{eq11} and including the terms coming from $d^2 E_{tot}^{(1)}/d u_{\mu s \alpha} d u_{\nu s' \beta}$, $d^2 E_{tot}^{(3)}/d u_{\mu s \alpha} 
d u_{\nu s' \beta}$, and $d^2 E_{tot}^{(4)}/d u_{\mu s \alpha} d u_{\nu s' \beta}$, defined in Ref. \onlinecite{us_fr_dfpt}.

\section{Applications}
In this section we use the theory described above to compute the phonon dispersions of ferromagnetic fcc Ni 
and of a monatomic ferromagnetic Pt nanowire. We validate the theory by comparing the phonon frequencies 
obtained by diagonalizing the dynamical matrix (Eq. \eqref{eq9}) with those obtained by the frozen phonon method.

\subsection*{Computational details}
First-principle calculations were performed within the Local Density Approximation (LDA) \cite{PZ} and the Perdew-Burke-Ernzerhof 
(PBE) \cite{PBE} schemes, as implemented in the Quantum ESPRESSO \cite{{QE},{QE_2}} and 
\texttt{thermo\_pw} \cite{thermo_pw} packages. The atoms are described by FR 
US PPs \cite{us_fr_pseudo}, with 4$s$ and 3$d$ valence electrons for Ni 
(PPs \texttt{Ni.rel-pz-n-rrkjus$\_$psl.0.1.UPF} and \texttt{Ni.rel-pbe-n-rrkjus$\_$psl.0.1.UPF} from pslibrary 0.1) and with 
6$s$ and 5$d$ valence electrons for Pt (\texttt{PP Pt.rel-pz-n-rrkjus$\_$psl.1.0.0.UPF} from pslibrary 1.0.0 
\cite{{pslibrary},{pslibrary_2}}).

DFPT calculations on ferromagnetic fcc Ni are at the theoretical LDA and PBE lattice constants, $a = 6.483$ a.u. 
and $a=6.658$ a.u., which are $2.6$\% and $0.02$\% smaller than experiment \cite{COD} ($a=6.659$ a.u.), 
respectively. The pseudo wavefunctions 
(charge density) are expanded in a plane waves basis set with a kinetic energy cut-off of $120$ ($600$) Ry. The 
BZ integrations were done using a shifted uniform Monkhorst-Pack \cite{k_grid} $\bm{k}$-point 
mesh of $28 \times 28 \times 28$ points for the phonon calculations at a single wave vector $\bm{q}$. 
The same computational parameters, except the $\bm{k}$-point mesh which has been reduced to $18 \times 18 \times 18$ 
points, have been used for the phonon dispersions. The dynamical matrices have 
been computed by DFPT on a $6 \times 6 \times 6$ $\bm{q}$-point mesh, and Fourier interpolated to obtain the 
complete dispersions. 
Phonon frequencies of ferromagnetic Ni with the frozen phonon method, were calculated with 
a simple cubic supercell with $4$ Ni atoms. The kinetic energy cut-offs used are the same as for the DFPT 
calculations, while the BZ integrations were performed on a $\bm{k}$-point mesh of $24 \times 24 \times 24$ points. 
The presence of a Fermi surface has been dealt with by the Methfessel-Paxton smearing method \cite{MP} with a 
smearing parameter $\sigma=0.02$ Ry. 

DFPT calculations on monatomic ferromagnetic Pt nanowire were done at a stretched geometry with interatomic distance 
$d = 4.927$ a.u.. The wire replicas have been separated by a vacuum space of $20$ a.u.. We have checked that 
by increasing the vacuum space the computed frequencies do not change more than $0.2$ cm$^{-1}$. The system has been 
studied in a ferromagnetic configuration, with magnetization either parallel or perpendicular to the wire. 
The kinetic energy cut-off was $60$ ($400$) Ry for the wave functions (charge density). The $\bm{k}$-point 
mesh is a shifted uniform Monkhorst-Pack mesh of $300$ points. Frozen phonon calculations 
were performed with supercells with $2$ and $4$ Pt atoms, and Monkhorst-Pack meshes of $150$ 
and $75$ $\bm{k}$-points, respectively. The smearing parameter was $\sigma = 0.002$ Ry.
 
\subsection*{Fcc Ni}
We start our discussion from the computation of the phonon frequencies of ferromagnetic fcc Ni 
with the magnetization along $[001]$ (and with a magnitude that turns out to be $0.62 \, \mu_B$ per atom), 
and compare the DFPT and the frozen phonon method at the $Y$ and $Z$ points. The results obtained are reported 
in Table \ref{t1}. The frequencies of the transverse modes at $\bm{q} = (0, 0, 2 \pi / a)$ (Z) are degenerate with 
both methods, as a consequence of the tetragonal magnetic symmetry ($D_{4h} (C_{4h})$): indeed both transverse modes 
have atomic displacements perpendicular to the magnetization. Instead, the transverse modes at 
$\bm{q} = (0, 2 \pi / a, 0)$ (Y) show a small splitting of $0.04 \, \text{cm}^{-1}$. The two modes are actually 
different because the atomic displacements are either parallel or perpendicular to the magnetization. A frequency 
splitting arises as a consequence of spin-orbit coupling. The DFPT and frozen phonon methods agree within 
$0.3 \, \text{cm}^{-1}$. The DFPT and the frozen phonon method predict the same splitting, which however is 
small compared to the agreement of the absolute values of the frequencies obtained with the two methods, hence 
it is not possible to give an accurate quantitative prediction, but only an order of magnitude. With the kinetic 
energy cut-offs and $\bm{k}$-point mesh used, the frequencies obtained are converged 
within $5 \times 10^{-3} \, \text{cm}^{-1}$, the same order of magnitude as the errorbar reported in Table \ref{t1}
and due to the fit.
\begin{table}[]
\centering
\begin{tabular}{lccc}
\hline
\hline
& DFPT  & & Frozen phonon \\
& $\nu (\text{cm}^{-1})$ & & $\nu (\text{cm}^{-1})$ \\
\hline
$T_{x}^{(0,1,0)}$ & $232.438$ & & $232.691 \pm 0.006$ \\
$T_{z}^{(0,1,0)}$ & $232.397$ & & $232.648 \pm 0.006$ \\
$T_{\{xy\}}^{(0,0,1)}$ & $232.433$ & & $232.688 \pm 0.006$ \\
\hline
\hline
\end{tabular}
\caption{Computed FR LDA phonon frequencies at $\bm{q} = (0, 2 \pi / a, 0)$ and $\bm{q} = (0, 0, 2 \pi / a)$ with DFPT and the frozen phonon method for fcc Ni. The magnetization is oriented along the $z$ axis. The subscripts indicate the polarization of the phonon modes.}
\label{t1}
\end{table}

\begin{figure}
\centering
\includegraphics[width=0.5\textwidth]{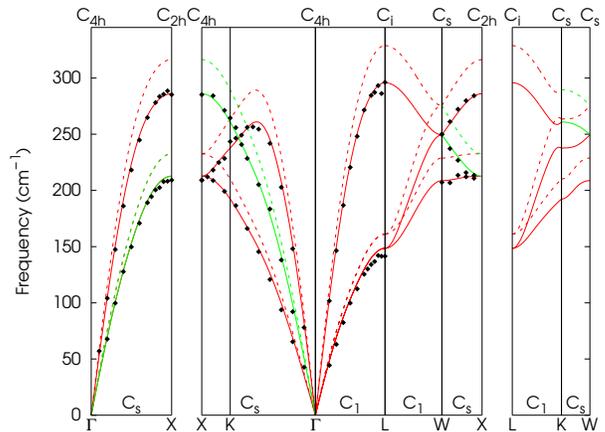}
\caption{Computed FR LDA (dashed lines) and PBE (solid lines) phonon dispersions of ferromagnetic fcc Ni, compared to inelastic
neutron scattering data (solid diamonds). Phonon modes are classified using symmetry, but only the operations that do not require $\m{T}$ are used.}
\label{f1}
\end{figure}

In Fig. \ref{f1} we show the complete phonon dispersion of fcc Ni obtained by DFPT. Both LDA and PBE theoretical 
dispersions are shown, together with inelastic neutron scattering data \cite{exp_phonon_Ni}. The agreement between 
the LDA result and the experiment is poor, mainly because LDA underestimates the lattice constant: the 
highest frequencies of the dispersion (e.g. at the $X$ and $L$ points) are about $30 \, \text{cm}^{-1}$ higher 
than the experiment. On the other hand, the PBE phonon dispersions are in excellent agreement with the experiment. 
Note however that this agreement is slightly worsened by temperature effects \cite{Ni_temp_phonon} not included 
in the present study.

\subsection*{Pt monatomic wire}
\begin{table*}[tp]
\centering
\begin{tabular}{c|l|ccc|ccc}
\hline 
\hline
q & & & $\bm{m} \parallel x$ & & & $\bm{m} \parallel z$ & \\
\hline
& & DFPT  & & Frozen phonon & DFPT  & & Frozen phonon\\
& & $\nu (\text{cm}^{-1})$ & & $\nu (\text{cm}^{-1})$ & $\nu (\text{cm}^{-1})$ & & $\nu (\text{cm}^{-1})$ \\
\hline
& $T_{x}$ & $36.51$  & & $37.02  \pm 0.03$ & $45.71$  & & $46.10 \pm 0.03$  \\  
$\pi/a$ & $T_{y}$ & $37.00$  & & $37.34  \pm 0.03$ & $45.71$  & & $46.10 \pm 0.03$  \\  
& $L$     & $113.98$ & & $114.21 \pm 0.03$ & $110.30$ & & $110.51 \pm 0.03$ \\
\hline
& $T_{x}$ & $25.1$ & & $25.5 \pm 0.1$ & $39.17$ & & $39.63 \pm 0.03$ \\  
$\pi/2a$ & $T_{y}$ & $32.1$ & & $31.8 \pm 0.1$ & $39.17$ & & $39.65 \pm 0.03$ \\  
& $L$     & $54.2$ & & $53.8 \pm 0.1$ & $62.66$ & & $63.14 \pm 0.03$ \\
\hline
\hline
\end{tabular}
\caption{Computed FR LDA phonon frequencies at $q = \pi/a$ and $q = \pi / 2a$ with DFPT and the frozen phonon method for a monatomic ferromagnetic Pt nanowire. The nanowire is oriented along the $z$ axis. Results are shown with both $\bm{m} \parallel x$ and $\bm{m} \parallel z$. The subscripts indicate the polarization of the phonon modes.}
\label{t2}
\end{table*}

In this section we consider a monatomic Pt nanowire, a metal with ferromagnetic ordering. It has been 
shown \cite{{Pt_nature},{Pt_PRB}} that at its equilibrium geometry (atomic distance $d = 4.441$ a.u.) 
the system shows a colossal magnetic anisotropy, since the preferred orientation of the magnetization is 
parallel to the wire and the magnetization vanishes when forced to be perpendicular to the wire. Instead, 
for stretched geometries with atomic distance higher than $4.913$ a.u. a non-zero magnetization perpendicular 
to the wire is allowed. Here we consider a stretched geometry with $d = 4.927$ a.u. and compute the phonon 
dispersions with both a magnetization parallel and perpendicular to the wire.
In the following the nanowire is along the $z$ direction. 
In Table \ref{t2} we compare the phonon frequencies, at $q = \pi/a$ and $q = \pi/2a$ with 
$\bm{m} \parallel x$ and $\bm{m} \parallel z$, computed by the DFPT and with the frozen phonon method. 
With a magnetization $\bm{m} \parallel z$ ($m = 0.65 \, \mu_B$ per atom), the frequencies of the transverse modes 
are degenerate, while with $\bm{m} \parallel x$ ($m = 0.13 \, \mu_B$ per atom) at $q = \pi/a$ the two transverse 
modes show a splitting of about $0.5 \, \text{cm}^{-1}$, which is of the same order of magnitude 
as the overall agreement of the two methods. At $q = \pi/2a$ this splitting is about 
$7 \, \text{cm}^{-1}$, one order of magnitude larger than at $q = \pi/a$. In both cases the polarization of 
the transverse mode with higher frequency is perpendicular to the magnetization. As discussed above for fcc Ni, 
the two transverse modes are not equivalent due to the 
presence of the magnetization and of spin-orbit coupling. Pt atoms are heavier than Ni and 
show a stronger spin-orbit interaction: indeed, the splitting reported for Pt is $1-2$ orders of magnitudes 
higher than in Ni. The DFPT and frozen phonon results agree within $0.4 \, \text{cm}^{-1}$ on average. 
As before, the errorbars reported in Table \ref{t2} come from the linear fit. With the kinetic energy cut-offs 
and the $\bm{k}$-point mesh used all the frequencies reported are converged within $0.03 \, \text{cm}^{-1}$ 

\begin{figure}
\centering
\includegraphics[width=0.5\textwidth]{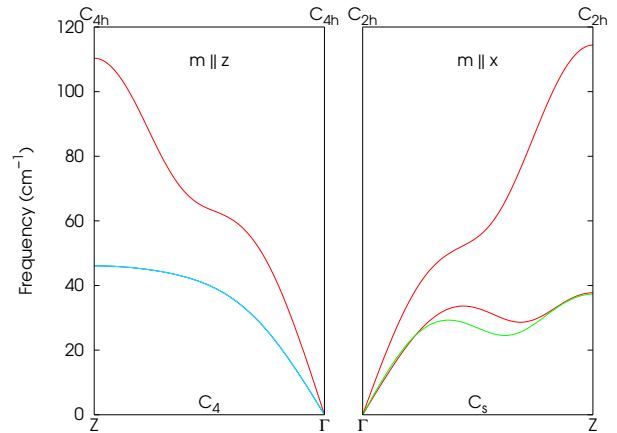}
\caption{Computed FR LDA phonon dispersions of ferromagnetic Pt nanowire. Left panel: 
magnetization parallel to the wire. Right panel: magnetization perpendicular to the wire.}
\label{f2}
\end{figure}

In Fig. \ref{f2} we show the phonon branches along $\Gamma - Z$ for a ferromagnetic wire with magnetization 
parallel (left panel) or perpendicular to the wire (right panel). The two dispersions show evident differences: 
at $q = \pi/a$, the longitudinal mode for the wire with $\bm{m} \parallel z$ is lower in frequency than for the wire 
with $\bm{m} \parallel x$, while the transverse modes are higher in frequency. 
In the central part of the BZ, around $q = \pi/2a$, the longitudinal mode of the 
wire with $\bm{m} \parallel z$ has a higher frequency at the Kohn anomaly than the wire with $\bm{m} \parallel x$, 
while the transverse modes show a Kohn anomaly only for $\bm{m} \parallel x$. We remark that at the stretched 
geometry studied ($d = 4.927$ a.u.) the phonon modes are still stable, but the range of atomic distances at which 
both modes are stable is quite narrow. 

\section{Conclusions}
We extended the DFPT for lattice dynamics with FR NC 
and US PPs to the magnetic case. 
We validated the theory by comparing the DFPT to the frozen phonon method for ferromagnetic fcc Ni and for a 
monatomic ferromagnetic Pt nanowire. The agreement between the two methods is within $0.5 \, \text{cm}^{-1}$. For 
both systems, we computed by DFPT also the complete phonon dispersions and discussed their features, 
showing that magnetism together with spin-orbit coupling may lift the degeneracy of some phonon modes. 
For our systems these splittings range from $10^{-2} \, \text{cm}^{-1}$ (in Ni) to a few $\text{cm}^{-1}$ 
(in Pt nanowire). 
\section*{Acknowledgments}

Computational facilities have been provided by SISSA through its Linux Cluster and ITCS and by CINECA through 
the SISSA-CINECA 2018-2019 Agreement.

\end{document}